\documentclass[epj,twocolumn]{webofc}
\usepackage[varg]{txfonts}
\woctitle{Hadron Collider Physics symposium 2012}

\begin{document}

 \title{Top properties in $t\bar{t}$ events at CMS (includes mass)}

 \author{V. Adler
  \inst{1}
  \fnsep
  \thanks{Speaker, on behalf of the CMS Collaboration.}
  \thanks{\email{volker.adler@cern.ch}}
 }
 \institute{Universiteit Gent, Department of Physics and Astronomy, Proeftuinstraat 86, B-9000 Gent, Belgium}

 \abstract{Selected results from the following topics are presented:
  Measurements of several top quark properties are obtained from the CMS data collected in 2011 at a center-of-mass energy of 7\,TeV.
  The results include measurements of the top quark mass, the W helicity in top decays, the top quark charge, and of the $t\bar{t}$ spin correlation and the search for anomalous couplings.
 }

 \maketitle

 \section{Introduction}
  \label{sec-1}

  As the heaviest particle in the Standard Model, the top quark is a unique probe for the model's validity.
  Due to its very high mass $m_t$ it decays rapidly before hadronization and thus allows for precise measurements of its properties.
  Possible deviations from Standard Model predictions can give rise to new physics beyond the Standard Model (BSM) like e.g. anomalous couplings or the existence of Supersymmetry (SUSY).

  The Large Hadron Collider (LHC) at CERN \cite{Evans:2008zzb} collides protons at center-of-mass energies of several TeV, which increases the production rate of heavy particles significantly compared to former colliders.
  In particular, the cross sections for top quark pair production \cite{Kidonakis:2010dk} together with the collision rate allow to refer to the LHC as ``top quark factory''.

  The CMS detector \cite{Chatrchyan:2008aa} is a multi-purpose detector at the LHC.
  All analyses described here have been performed using data collected with the CMS detector during the 2011 data taking period with proton-proton collisions at a center-of-mass energy of 7\,TeV, which amounts to an integrated luminosity of 5\,fb$^{-1}$ in total.

 \section{The Mass of the Top Quark}
  \label{sec-2}

   Measurements of $m_t$ are performed using various decay channels of $t\bar{t}$ pairs with different analysis techniques, which are presented in Sects.~\ref{sec-2-1}--\ref{sec-2-4}, followed by the CMS top quark mass combination in Sect.~\ref{sec-2-5}.
   A measurement of the mass difference between top and anti-top quarks is described in Sect.~\ref{sec-2-6}.

  \subsection{Measurement in $t\bar{t}$ Events with Dilepton Final States}
   \label{sec-2-1}

   A measurement of $m_t$ is performed with events in the dileptonic decay channels $ee$, $\mu\mu$ and $e\mu$ \cite{Chatrchyan:2012ea}.
   Candidate $t\bar{t}$ events are selected with two isolated, oppositely charged leptons with high transverse momentum $p_t$ and two or more jets, where at least one has to be identified as originating from a $b$-quark ($b$-tagged).

   Including constraints on the $W$-boson mass and the equality of the masses of top and anti-top quark, still one parameter in the reconstruction of $t\bar{t}$ events remains free in these decay channels.
   This parameter needs to be constrained by imposing an additional hypothesis.
   In this analysis, the mass of the top quarks is reconstructed using an improved Matrix Weighting Technique \cite{Abbott:1997fv}, the Analytical Matrix Weighting Technique (AMWT).
   Instead of simply using the two leading jets in the reconstruction, leading $b$-tagged jets are considered first and supplemented by the leading untagged jets then if necessary.
   This increases the fraction of correct jet assignments significantly.
   As additional constraint a hypothesis on $m_t$ itself is used.
   The reconstructed mass $m_{AMWT}$ per event is then the hypothetical mass giving the highest event weight, which is computed from a probability density and takes resolution effects into account.

   Finally, likelihoods for values of $m_t$ between 161.5 and 184.5\,GeV are computed from the distributions of $m_{AMWT}$, and the mass value maximizing the likelihood is fitted.
   The top quark mass is measured to be
   $$m_t = 172.5 \pm 0.4\,\text{(stat.)} \pm 1.5\,\text{(syst.)}\,\text{GeV}$$
   with the dominant systematic effects coming from the jet energy scales (JES).
   This is the most precise measurement of $m_t$ in the dilepton channel to date.

  \subsection{Measurement in $t\bar{t}$ Events with Lepton+Jets Final States}
   \label{sec-2-2}

   The presented analysis measures $m_t$ in the semileptonic decay channels with an electron or muon in the final state \cite{Chatrchyan:2012cz}.
   Events with exactly one isolated lepton with high $p_t$ and at least four or more jets are selected, where at least two have to be $b$-tagged.

\begin{figure}
    \centering
    \includegraphics[width=0.45\linewidth,clip]{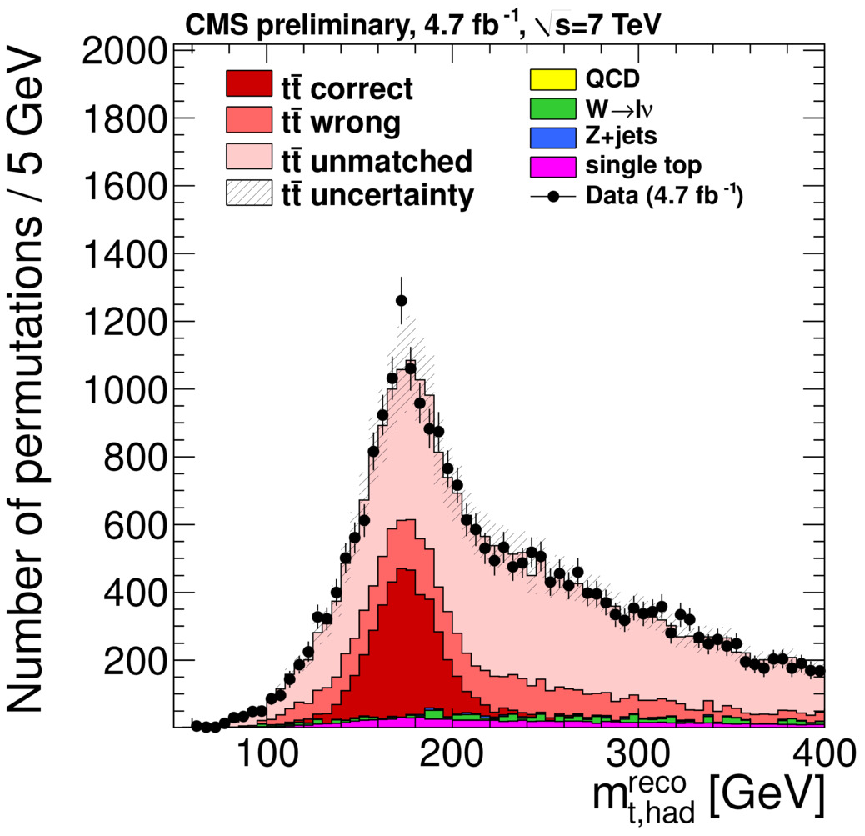}
    \hfill
    \includegraphics[width=0.45\linewidth,clip]{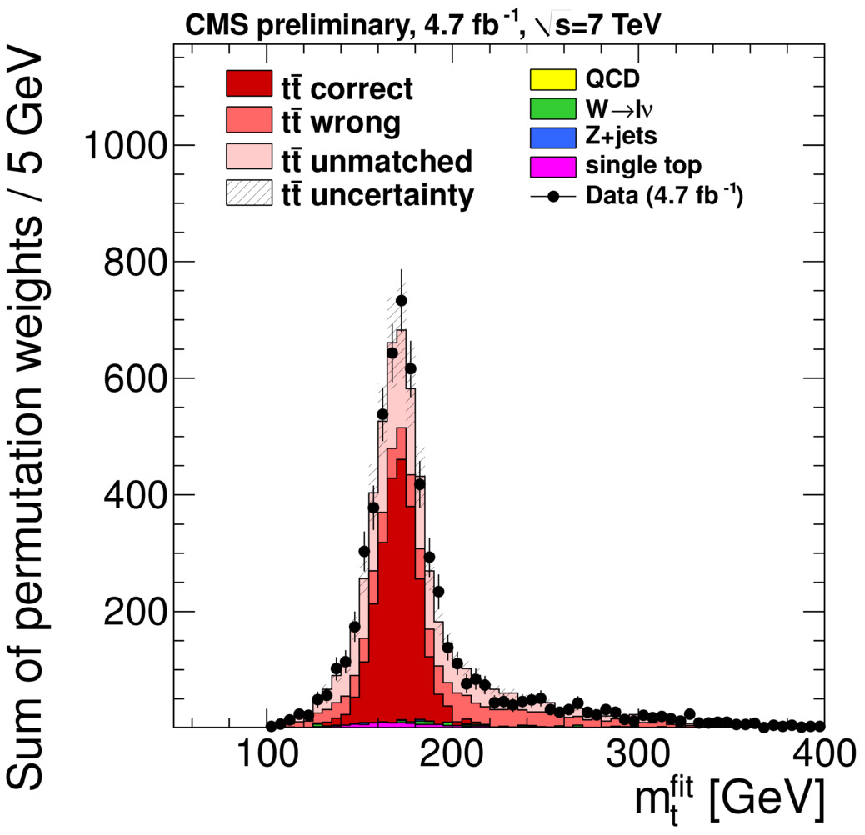}
    \caption{Reconstructed top quark masses, prior to kinematic fitting {\it (left)} and after fitting, goodness-of-fit selection and weighting by $P_{gof}$ {\it (right)}.
     Top quarks are simulated with $m_t = 172.5\,\text{GeV}$.}
    \label{fig-2-2-1}
   \end{figure}

   Using constraints on the $W$-boson mass and the equality of the masses of top and anti-top quark, it is generally possible to fully reconstruct the $t\bar{t}$ system for each event.
   This is done here by a kinematic fit, which is specialized for the semileptonic event topology \cite{Abbott:1998dc}.
   The measurement is then performed using the ideogram method, where the results from the kinematic fits of all possible jet combinations are considered, weighted by the goodness-of-fit probability $P_{gof}$.
   Only fit results with $P_{gof} \ge 0.2$ are taken into account.
   The effect of the kinematic fit and the weighting is illustrated in Fig.~\ref{fig-2-2-1}.

\begin{figure}
    \centering
    \sidecaption
    \includegraphics[width=0.5\linewidth,clip]{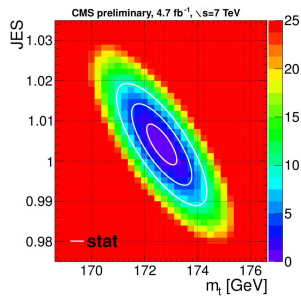}
    \caption{The 2D likelihood ($-2\delta\;log(\mathcal{L})$) measured for the $l$+jets final state.
     The ellipses correspond to statistical uncertainties on $m_t$ and JES of one, two and three standard deviations.}
    \label{fig-2-2-2}
   \end{figure}

   Likelihoods for the reconstructed $W$-boson and top quark masses are evaluated from analytic expressions obtained using simulation.
   This is done in a two-dimensional (2D) grid of hypotheses for $m_t$ and JES, which provides a simultaneous measurement of the JES for {\em light} jets from the well known mass of the $W$-boson.
   The 2D likelihood distribution is show in Fig.~\ref{fig-2-2-2}.
   Its minimization yields a top quark mass of
   $$m_t = 173.49 \pm 0.43\,\text{(stat.+JES)} \pm 0.98\,\text{(syst.)}\,\text{GeV}$$
   and a JES of
   $$JES = 0.994 \pm 0.003\,\text{(stat.)} \pm 0.008\,\text{(syst.).}$$
   This is the most precise single measurement of the top quark mass to date.
   After the elimination of the JES for light jets as dominant systematic uncertainty, the JES for $b$-jets and color reconnection effects become most important.

  \subsection{Measurement in $t\bar{t}$ Events with All-Jets Final States}
   \label{sec-2-3}

   The measurement of $m_t$ in the all-hadronic final state \cite{CMS:2012qxa} uses a similar combination of kinematic fit and ideogram method as in the semi-leptonic channel in Sect.~\ref{sec-2-2}.
   Events with at least six jets are selected, where at least two jets are required to be $b$-tagged.
   Compared to the semi-leptonic channel, a different fitting technique is utilized \cite{D'Hondt:2006bc}, and jet combinations with $P_{gof} \ge 0.09$ are accepted.
   In this case, the simultaneous determination of the JES does not improve the precision of the measurement due to the increased influence of other effects.
   Using a JES fixed at unity, the one-dimensional (1D) likelihood is minimized for a top quark mass of
   $$m_t = 173.49 \pm 0.69\,\text{(stat.)} \pm 1.25\,\text{(syst.)}\,\text{GeV,}$$
   with the JES as dominant systematic uncertainty.

  \subsection{Measurement in $t\bar{t}$ Events with Kinematic Endpoints}
   \label{sec-2-4}

   In the presence of invisible particles in a decay chain, the mass of the decaying particle cannot be reconstructed directly, but assumptions have to be added as constraints to the kinematic equations to substitute the missing observables.
   However, using only the visible kinematics in the reconstruction still would result in the correct mass in the extreme case of the invisible particles carrying zero momentum.
   At collider experiments, transverse mass distributions can be extrapolated to this extreme case, the kinematic endpoint.

\begin{figure}
    \centering
    \includegraphics[width=\linewidth,clip]{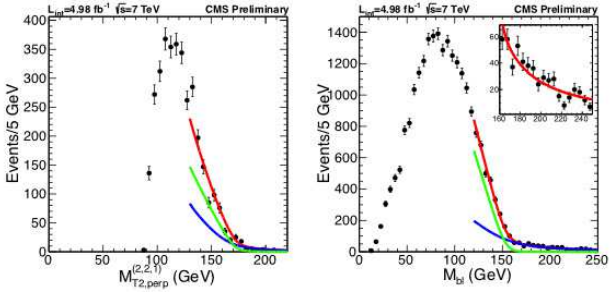}
    \caption{Doubly constrained simultaneous fits of $m_t$, using one $M_{T2\bot}$ {\it (left)} and $M_{bl}$ {\it (right)}.
     The red lines are the full fit, and the blue and green curves represent the background and signal shapes, respectively.
     The inset shows a zoom of the tail region of the right distribution.}
    \label{fig-2-4-1}
   \end{figure}

   The analysis presented here is performed in the dilepton channel \cite{CMS:2012eya}, selecting events with two isolated, oppositely charged leptons with high $p_T$, two $b$-tagged jets and high missing transverse energy $E_T^{miss}$.
   Two specially designed transverse mass expressions $M_{T2\bot}$ and the invariant mass of a lepton and a $b$-jet $M_{bl}$ are evaluated, where the endpoints of the distributions depend on $m_t$.
   An unbinned likelihood fit to the endpoint regions of the corresponding distributions in data with predefined signal and background shapes determines the mass value.
   The best precision is achieved with the neutrino and $W$-boson masses constrained, where only one of the $M_{T2\bot}$ is used.
   Fig.~\ref{fig-2-4-1} shows the fitted distributions.
   With this technique, the top quark mass is measured as
   $$m_t = 173.9 \pm 0.9\,\text{(stat.)} ^{+1.2}_{-1.8}\,\text{(syst.)}\,\text{GeV.}$$

  \subsection{Mass Combination}
   \label{sec-2-5}

\begin{figure}
    \centering
    \includegraphics[width=\linewidth,clip]{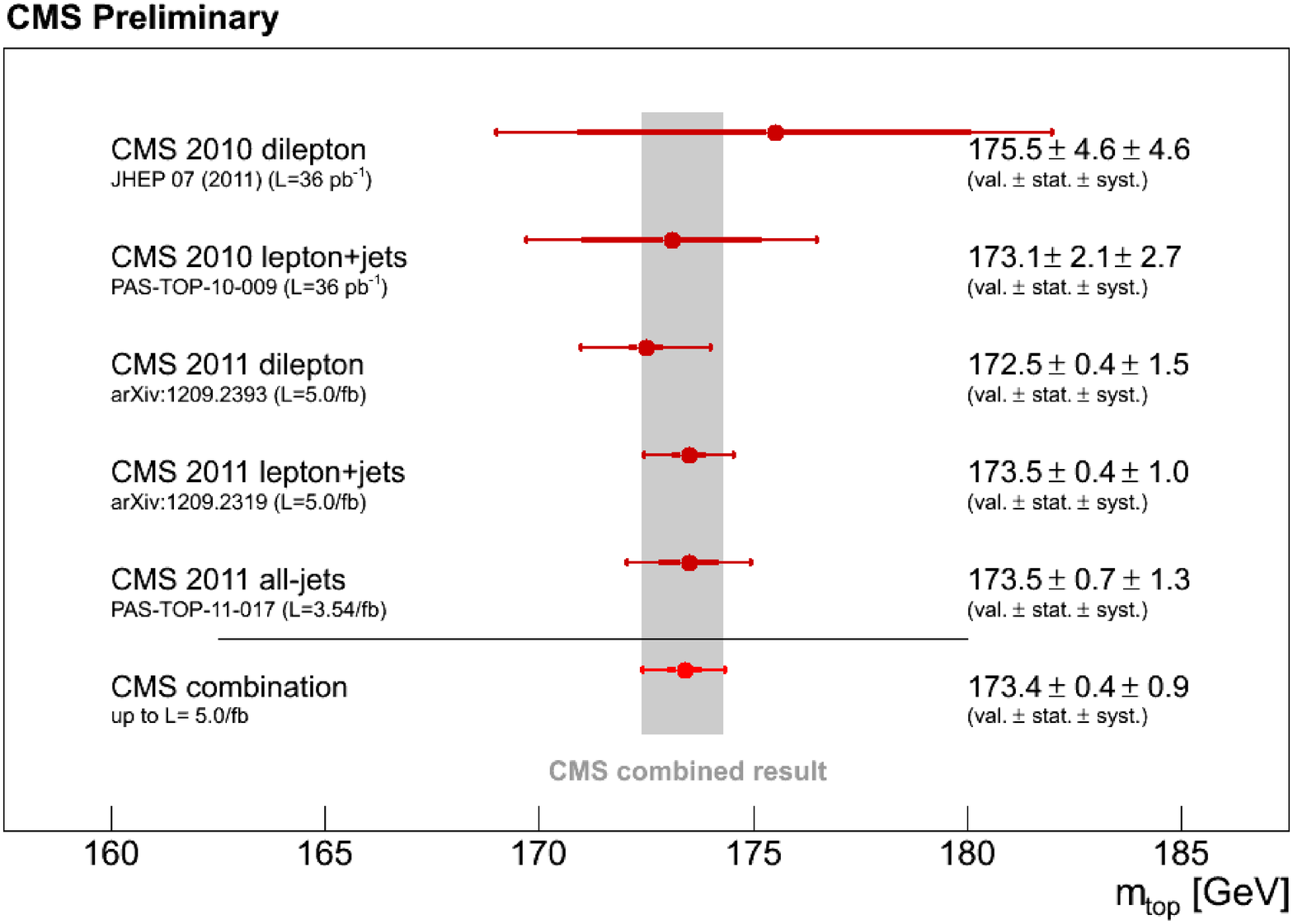}
    \caption{Summary of CMS $m_t$ measurements and their combination.}
    \label{fig-2-5-1}
   \end{figure}

   The combination of various measurements of $m_t$ presented here \cite{CMS:2012fya} uses the measurements described in Sects.~\ref{sec-2-1}--\ref{sec-2-3} and measurements performed with data collected in 2010 \cite{Chatrchyan:2011nb}.
   The individual measurements are combined using the Best Linear Unbiased Estimator (BLUE) method \cite{Lyons:1988rp}.
   The combination yields
   $$m_t = 173.36 \pm 0.38\,\text{(stat.)} \pm 0.91\,\text{(syst.)}\,\text{GeV.}$$
   Fig.~\ref{fig-2-5-1} summarizes the individual measurements and their combination.

  \subsection{Measurement of the Mass Difference between Top and Anti-Top Quarks}
   \label{sec-2-6}

   In order to determine a possible mass difference between top and anti-top quark $\Delta m_t = m_t - m_{\bar{t}}$, semileptonic $t\bar{t}$ events are selected similarly to the $m_t$ measurement described in Sect.~\ref{sec-2-2} \cite{Chatrchyan:2012uba}.
   A kinematic fit constrained by the $W$-boson mass is applied only to the hadronic part of each jet combination, and the masses are determined with the ideogram method, but separately for each reconstructed lepton charge.
   Masses from events with positively (negatively) charged lepton are assigned to the top (anti-top) quark.
   The difference between the two mass values is found to be
   $$\Delta m_t = -0.44 \pm 0.46\,\text{(stat.)} \pm 0.27\,\text{(syst.),}$$
   which is consistent with 0.
   The systematic uncertainty remains small, since many significant uncertainties cancel in the measurement of a mass {\em difference}.

 \section{Properties of the Top Quark}
  \label{sec-3}

  \subsection{First Measurement of $\mathcal{B}(t \rightarrow Wb)/\mathcal{B}(t \rightarrow Wq)$ in the Dilepton Channel}
   \label{sec-3-1}

   In the Standard Model, the magnitude of the Cabibbo-Kobayashi-Maskawa (CKM) matrix element $|V_{tb}|$ is expected to be close to unity, so that the top quark should decay into a $W$-boson and a $b$-quark almost exclusively.
   A deviation from this prediction could be a hint for a fourth quark generation.

\begin{figure}
    \centering
    \sidecaption
    \includegraphics[width=0.5\linewidth,clip]{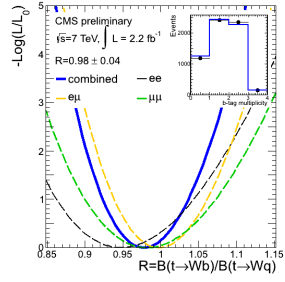}
    \caption{Variation of the profile likelihood used to extract $R$ from data.
     The variation observed in the exclusive dilepton channels is depicted by the dashed lines.
     The inset shows the inclusive $b$-tag multiplicity distribution and the result of the fit.}
    \label{fig-3-1-1}
   \end{figure}

   A measurement of the branching fraction ratio $R = \mathcal{B}(t \rightarrow Wb)/\mathcal{B}(t \rightarrow Wq)$ is performed in the dileptonic channel \cite{CMS:2012nua}.
   The number of expected $b$-tags per event is modeled as a function of $R$ for different event categories, and its most likely value is found after maximizing a binned likelihood function using that model and the observation from data.
%    The inclusive likelihood is shown in Fig.~\ref{fig-3-1-1} on the left.
   The inclusive likelihood is shown in Fig.~\ref{fig-3-1-1}.
   In this analysis, a branching fraction ratio of
   $$R_{obs} = 0.98 \pm 0.04\,\text{(stat.+syst.)}$$
   is observed, which is in agreement with Standard Model predictions.
   The systematic uncertainties are dominated by the $b$-tagging efficiency uncertainty and the uncertainty on the fraction of correct jet assignments.

   With the constraint $R \le 1$, a 95\% C.L. interval of $0.85 \le R \le 1.0$ is determined.

  \subsection{Search for Flavor Changing Neutral Currents in Top Quark Decays}
   \label{sec-3-2}

   The branching fraction for flavor changing neutral current (FCNC) decays of the top quark ($t \rightarrow Zq$) is expected to be extremely small.
   However, some BSM scenarios like e.g. R-parity violating SUSY predict measurable enhancements.

\begin{figure}
    \centering
    \includegraphics[width=0.45\linewidth,clip]{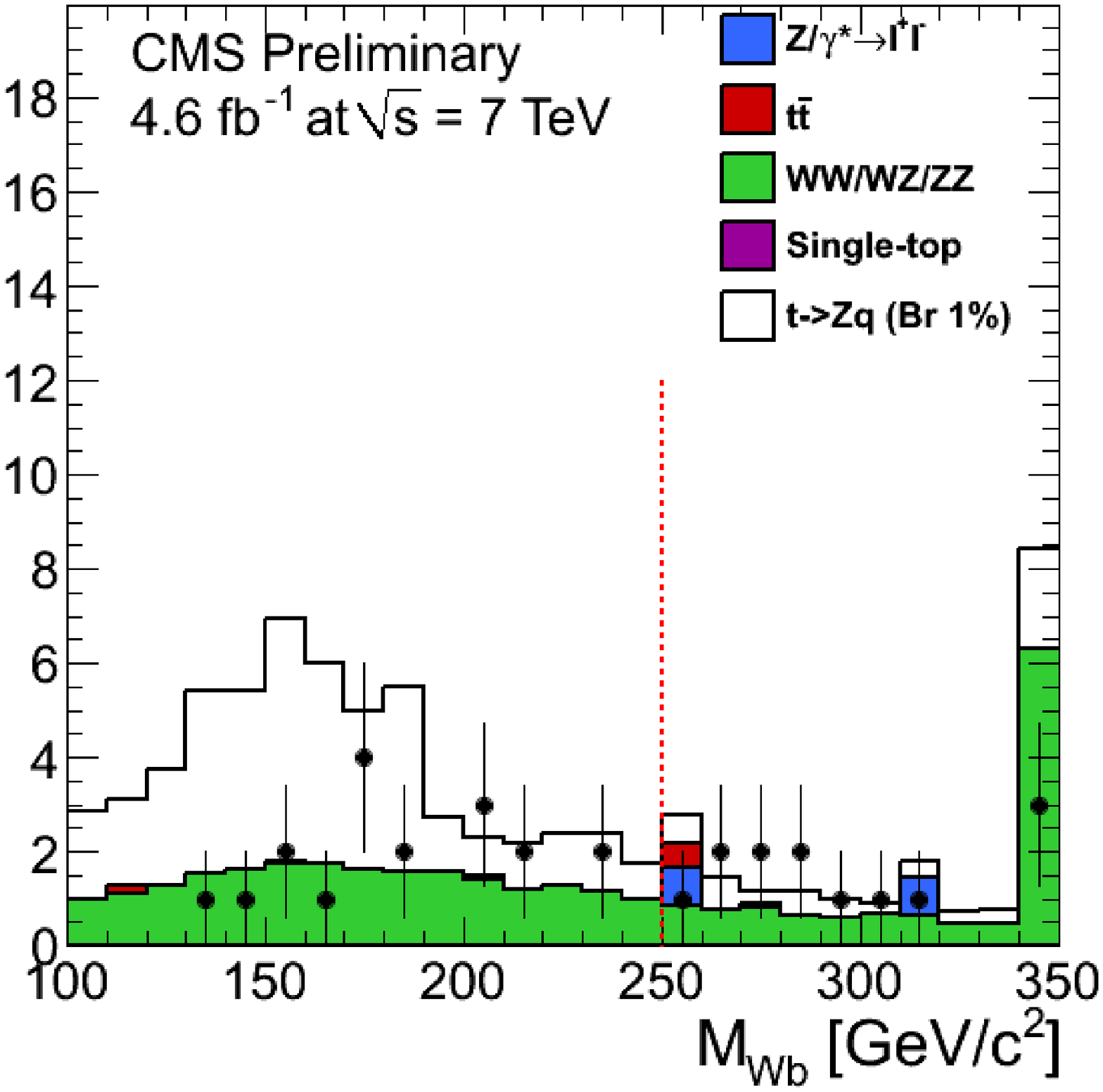}
    \hfill
    \includegraphics[width=0.45\linewidth,clip]{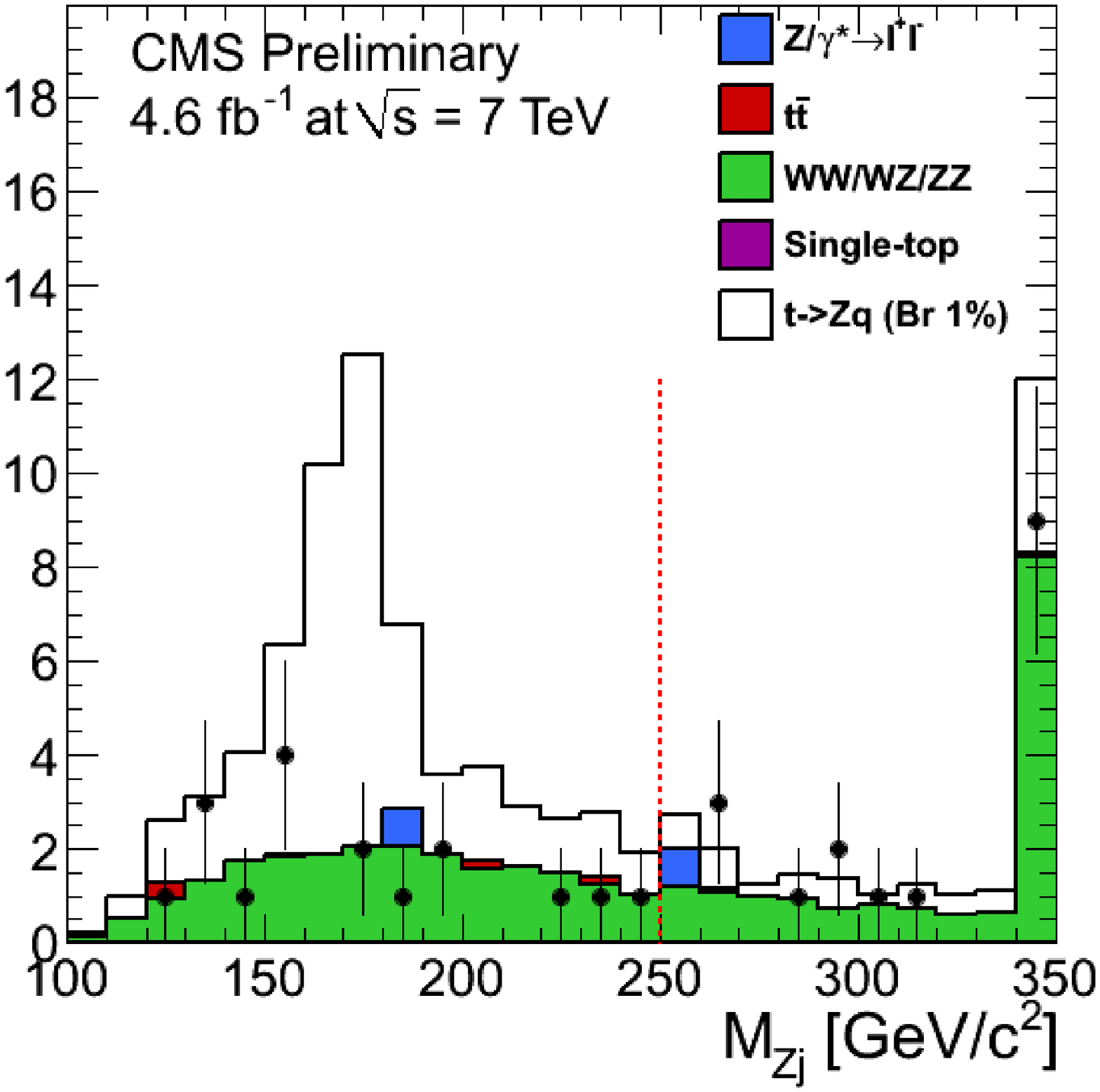}
    \caption{Comparison between data and simulation of the $m_{Wb}$ {\it (left)} and $m_{Zj}$ {\it (right)} distributions after the basic event selection.
     The open histogram shows the expected signal for $\mathcal{B}(t \rightarrow Zq) = 1\%$.
     The red dotted lines show the boundaries of the allowed mass region.}
    \label{fig-3-2-1}
   \end{figure}

   A search for FCNC in top quark decays is presented, which is performed using an event topology with three leptons from the decay chain $t\bar{t} \rightarrow Zq + Wb \rightarrow llq + l\nu b$, where $l$ can be $e$ or $\mu$ \cite{:2012sda}.
   Events with three isolated leptons with high $p_T$, at least two jets and a high $E_T^{miss}$ are selected.
   In addition, the scalar $p_T$ sum $S_T$ is required to be larger than 250\,GeV, and the reconstructed top quark masses have to be within 100 and 250\,GeV.
   The corresponding mass distributions are shown in Fig.~\ref{fig-3-2-1}.
   No deviation from the Standard Model is found and an observed upper limit at 95\% C.L. on $\mathcal{B}(t \rightarrow Zq)$ of 0.21\% is determined, where 0.40\% have been expected.
   A more stringent event selection using $b$-tags instead of $S_T$ confirms this result.

  \subsection{Constraints on the Top Quark Charge from $t\bar{t}$ Events}
   \label{sec-3-3}

   The Standard Model predicts a top quark charge of $+2/3e$.
   This analysis evaluates this prediction compared to an exotic top quark charge of $-4/3e$ \cite{CMS:2012oua}.

\begin{figure}
    \centering
    \includegraphics[width=0.45\linewidth,clip]{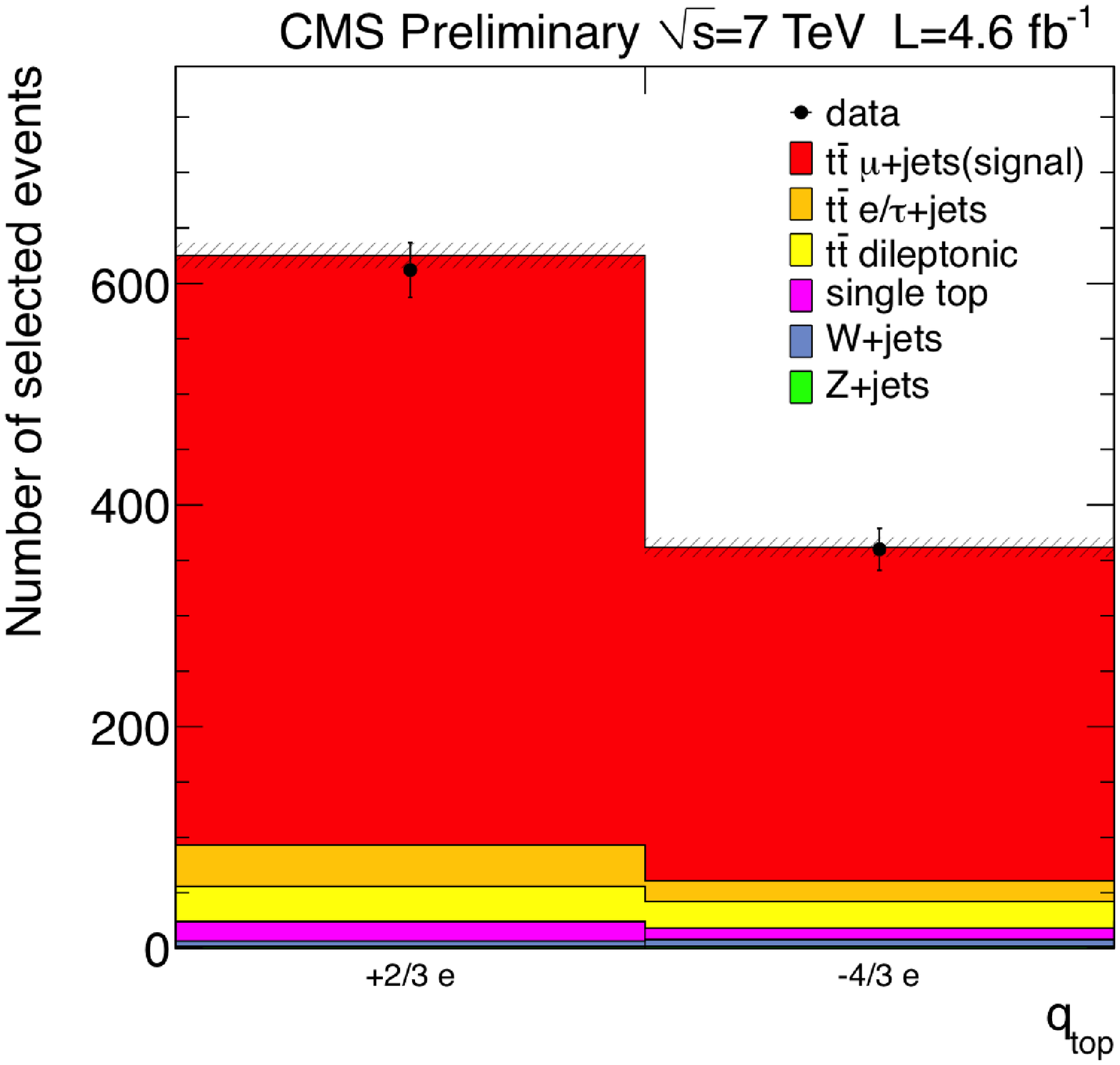}
    \hfill
    \includegraphics[width=0.45\linewidth,clip]{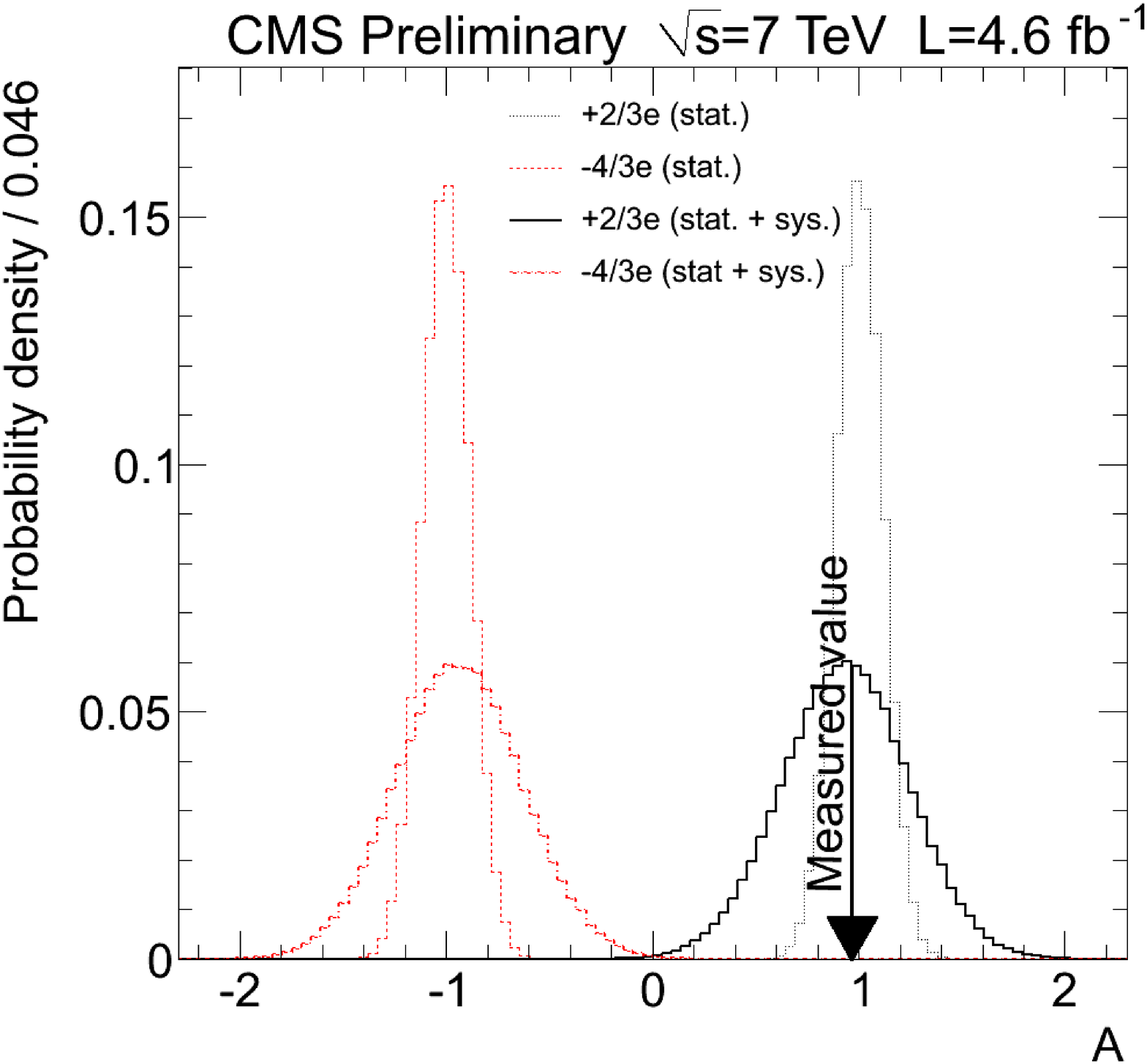}
    \caption{{\it Left}\,: Charge assigned to the top quark in the selected $t\bar{t}$ events.
     Data are compared to the Standard Model prediction with its statistical uncertainty represented by the shaded area.
     {\it Right}\,: Probability function of the asymmetry obtained from pseudo-experiments for the Standard Model and the exotic hypothesis.}
    \label{fig-3-3-1}
   \end{figure}

   Semileptonic events with muons are selected, with two of the four jets being $b$-tagged.
   Using the best approximation to the top quark mass, the tagged $b$-jets are assigned to either the leptonically or hadronically decaying top quark.
   In the presence of a soft muon in at least on of the $b$-jets, the charge of both $b$-quark can be determined.
   The charge of the leptonically decaying top quark is calculated as $q_t = q_{\mu} + q_b$, where $q_{\mu}$ is the reconstructed charge of the isolated muon and $q_b$ the charge determined for the $b$-jet assigned to this decay branch.
   The resulting charge distribution is shown in Fig.~\ref{fig-3-3-1} on the left.

   The agreement with the Standard Model is tested with a normalized asymmetry, which is expected to be +1(-1) for the Standard Model (exotic model) hypothesis.
   The measured asymmetry
   $$A = 0.97 \pm 0.12\,\text{(stat.)} \pm 0.31\,\text{(syst.)}$$
   is in agreement with the Standard Model.
   The measurement is compared to pseudo-experiments in Fig.~\ref{fig-3-3-1} on the right.

  \subsection{$W$ Helicity in $t\bar{t}$ Events}
   \label{sec-3-4}

   The Standard Model predicts helicity fractions for left-handed, right-handed and longitudinal $W$-bosons from top quark decays, $F_L$, $F_R$ and $F_0$.
   Deviations from these predictions can be interpreted in terms of anomalous $Wtb$ couplings.

\begin{figure}
    \centering
    \includegraphics[width=0.45\linewidth,clip]{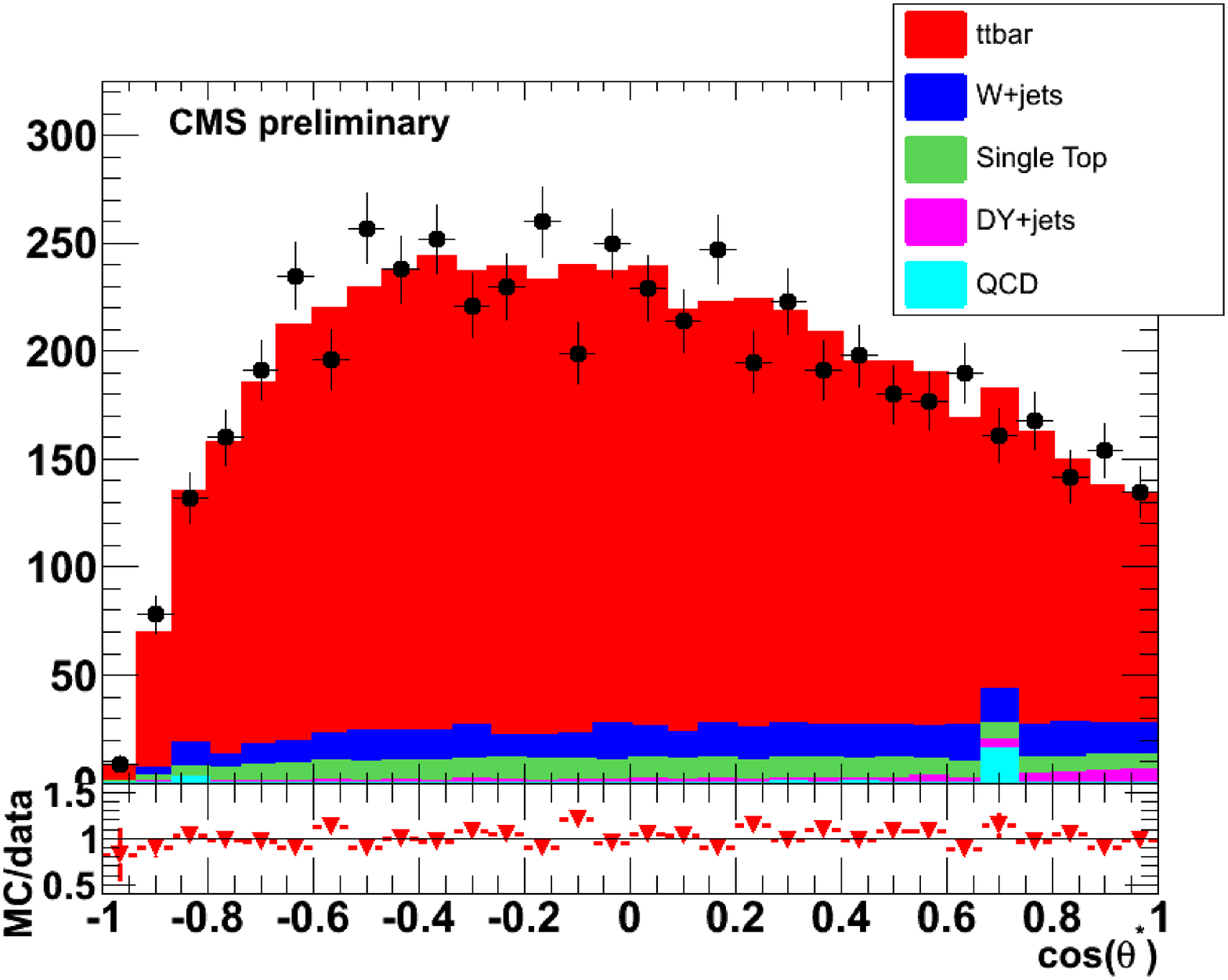}
    \hfill
    \includegraphics[width=0.45\linewidth,clip]{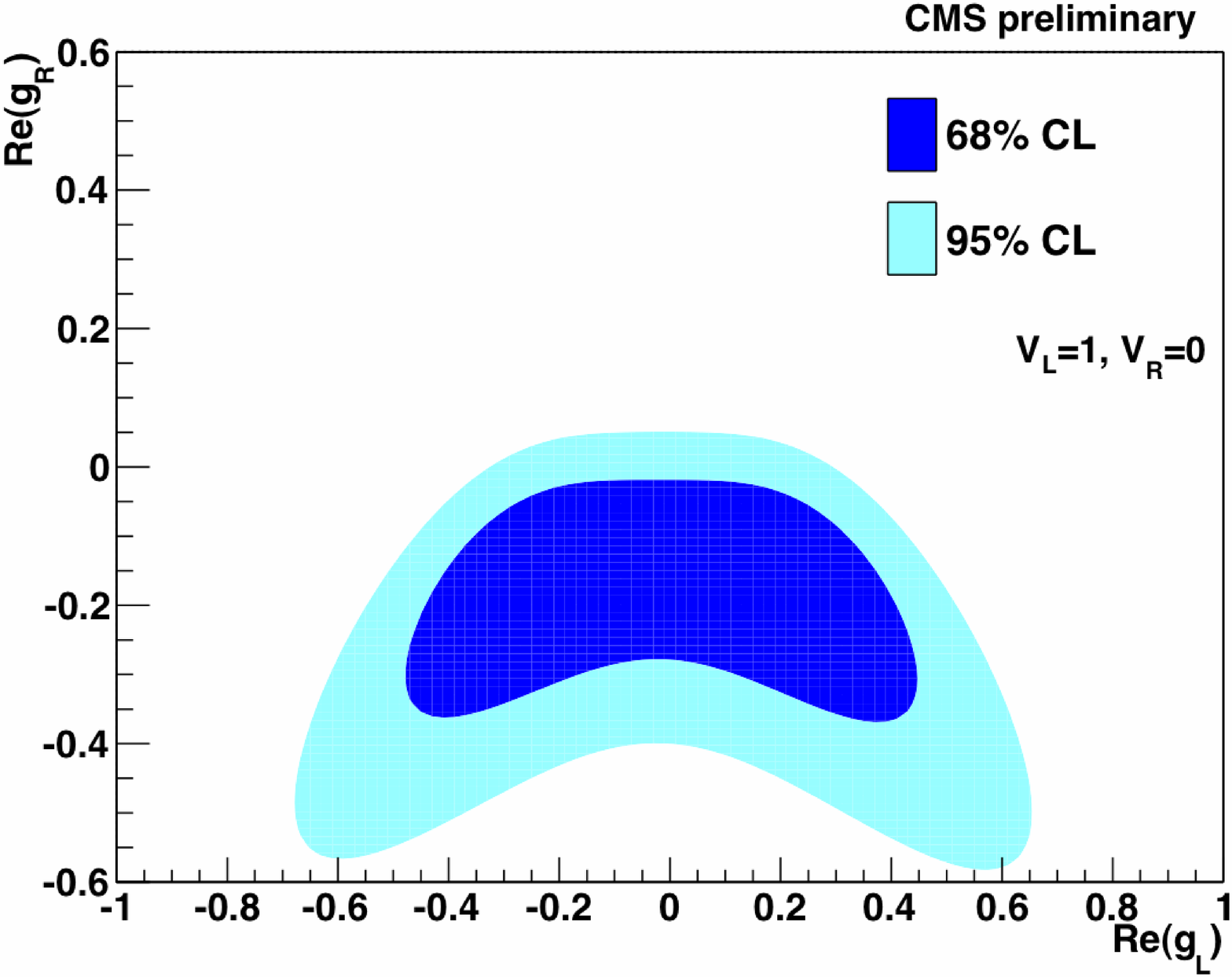}
    \caption{{\it Left}\,: The $cos(\theta^{\ast})$ distribution, from where the $W$ helicity fractions are obtained.
     The $t\bar{t}$ sample is additionally scaled by a factor of about 10\%.
     {\it Right}\,: Limits on the real components of the anomalous couplings $g_L$ and $g_R$.}
    \label{fig-3-4-1}
   \end{figure}

   A measurement of the helicity fractions is performed in the semileptonic decay channel with muons \cite{CMS:2012tta}.
   The $t\bar{t}$ system is fully reconstructed with a kinematic fit and the helicity angle $\theta^{\ast}$ is extracted.
   This angle is defined as the angle between the muon three-momentum in the rest frame of its parent $W$-boson and the $W$-boson's three-momentum in the rest frame of its parent top quark.
   The resulting distribution is shown in Fig.~\ref{fig-3-4-1} on the left.

   The dependency
   $$\frac{1}{\Gamma}\frac{d\Gamma}{d cos \theta^{\ast}} = \frac{3}{8}\left( 1 - cos \theta^{\ast} \right)^2 F_L + \frac{3}{8}\left( 1 + cos \theta^{\ast} \right)^2 F_R + \frac{3}{4}sin^2 \theta^{\ast} F_0$$
   allows to determine the helicity fractions by minimizing a Poisson likelihood.
   All results are in agreement with the Standard Model, and limits on the anomalous coupling constants $g_L$ and $g_R$ can be set, as illustrated in Fig.~\ref{fig-3-4-1} on the right.

  \subsection{Measurement of Spin Correlations in $t\bar{t}$ Production}
   \label{sec-3-5}

   Since the top quark decays before hadronization, its spin is transferred to its decay products and thus become experimentally accessible.
   Measuring the spin correlations in $t\bar{t}$ events probes perturbative QCD at the production vertex.

   The analysis presented here uses events in the dileptonic decay channels $ee$, $\mu\mu$ and $e\mu$ \cite{CMS:2012cxa}.
   In this topology, the angular difference in the azimuthal plane between the two leptons $\Delta \phi_{l^+ l^-} = | \phi_{l^+} - \phi_{l^-} |$ is sensitive to the spin correlations.

\begin{figure}
    \centering
    \includegraphics[width=0.7\linewidth,clip]{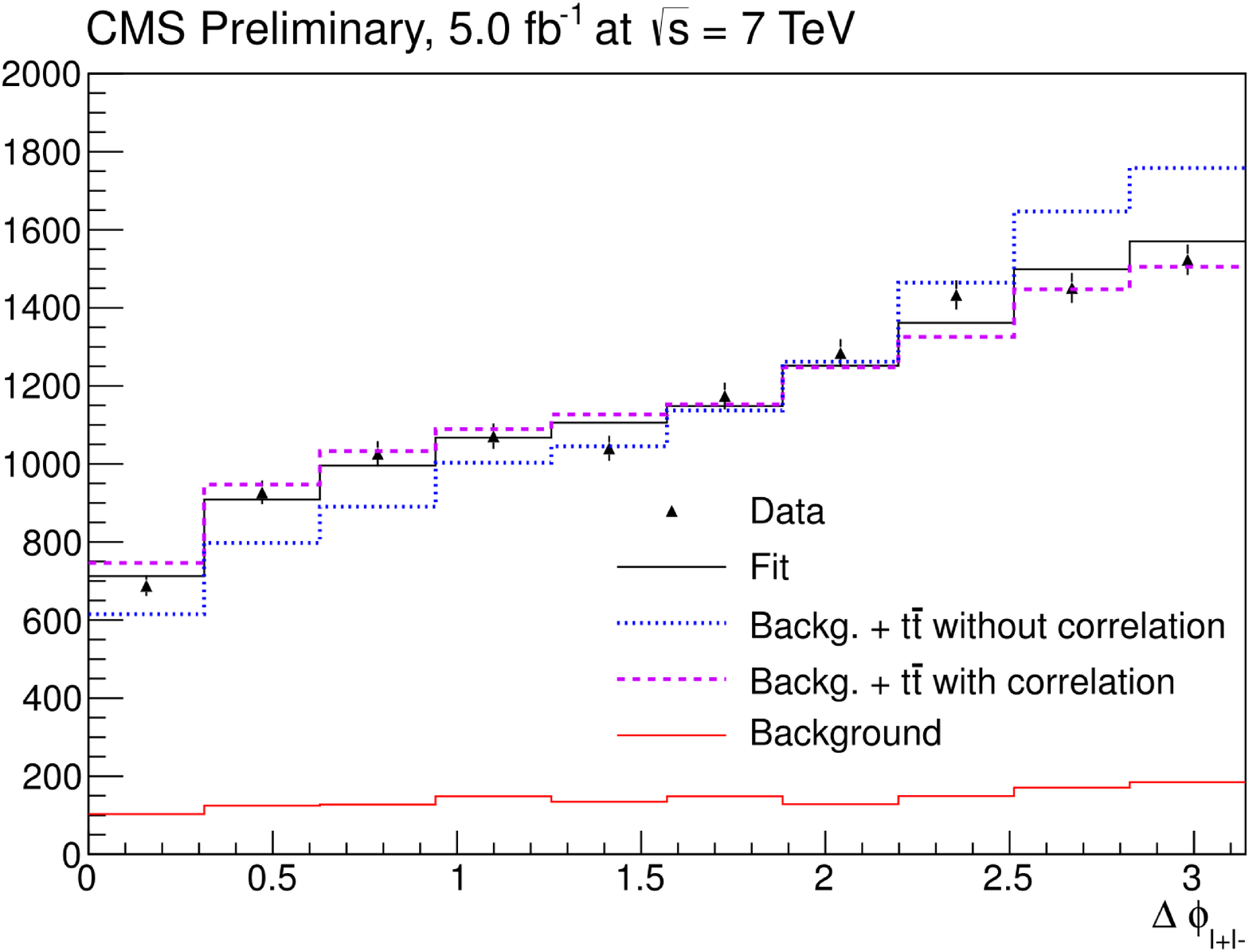}
    \caption{Result of the fit performed on data after the combination of the three channels.
     The hypotheses with and without spin correlations are also shown, including the background component.}
    \label{fig-3-5-1}
   \end{figure}

   The spin correlation coefficient is determined by a simultaneous template fit to the $\Delta \phi_{l^+ l^-}$ distributions in each decay mode.
   Templates are considered for $t\bar{t}$ events with and without spin correlations at next-to-leading order (NLO) and background events, individually for each channel.
   The combined fit result is depicted in Fig.~\ref{fig-3-5-1}.
   The measurement obtains a spin correlation coefficient in the helicity basis of
   $$A_{hel}^{meas} = 0.24 \pm 0.02\,\text{(stat.)} \pm 0.08\,\text{(syst.),}$$
   which is in agreement with the Standard Model expectation of $A_{hel}^{SM} = 0.31$ from NLO calculations.
   This agreement is confirmed by further measurements with asymmetry distributions, inclusively and at a high invariant mass of the $t\bar{t}$ system, and inclusively unfolded to parton level.

  \subsection{Measurement of the Top Quark Polarization in the Dilepton Final State}
   \label{sec-3-6}

\begin{figure}
    \centering
    \sidecaption
    \includegraphics[width=0.5\linewidth,clip]{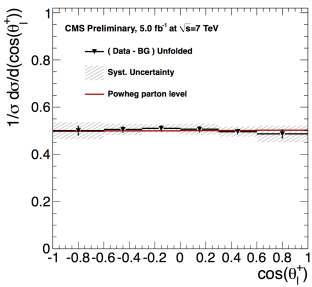}
    \caption{Background-subtracted and unfolded $cos(\theta)^{+}_l$ distribution compared to the parton level distribution from the used Monte Carlo generator.
     The error bars represent the statistical uncertainties.
     The bin values are correlated due to unfolding.}
    \label{fig-3-6-1}
   \end{figure}

   The dileptonic final states with $ee$, $\mu\mu$ and $e\mu$ are also used to measure the polarization of the top quark \cite{CMS:2012owa}.
   The top quark's polarization along a given axis $\hat{n}$ can be extracted from the asymmetry
   $$P_n = \frac{N( cos( \theta_l^+ ) > 0 ) - N( cos( \theta_l^+ ) < 0 )}{N( cos( \theta_l^+ ) > 0 ) + N( cos( \theta_l^+ ) < 0 )}\text{,}$$
   where $\theta_l^+$ is the production angle of the positively charged lepton in the rest frame of its parent top quark with respect to the direction of the parent top quark in the rest frame of the $t\bar{t}$ system.
   In this measurement, the background-subtracted data distribution is corrected to parton level using a regularized unfolding procedure.
   The result of the unfolding is shown in Fig.~\ref{fig-3-6-1}.
   The extracted polarization of
   $$P_n = -0.009 \pm 0.029\,\text{(stat.)} \pm 0.041\,\text{(syst.)}$$
   is in agreement with the Standard Model prediction as implemented in the used Monte Carlo event generator.

 \section{Conclusions}
  \label{sec-4}

  The high production rate of top quarks at the LHC allows for precision measurements of the quark's properties.
  The CMS experiment has performed measurements of top quark properties, using data from proton-proton collisions at a center-of-mass energy of 7\,TeV.

  The top quark mass is measured to high precision as
  $$m_t = 173.36 \pm 0.38\,\text{(stat.)} \pm 0.91\,\text{(syst.),}$$
  and the mass difference between top and anti-top quark is compatible with 0.

  All measurements of couplings, charge or spin related properties are in agreement with Standard Model predictions.

\end{document}